\documentclass[preprint,12pt]{elsarticle}

\usepackage{amssymb}
\usepackage{amsmath}

\usepackage{cleveref}

\usepackage{url}

\usepackage{enumitem}

\usepackage{booktabs}
\usepackage{array}
\newcolumntype{P}[1]{>{\centering\arraybackslash}p{#1}}

\begin{document}

\begin{frontmatter}



\title{Spectral and Rhythm Feature Performance Evaluation for Category and Class Level Audio Classification with Deep Convolutional Neural Networks}


\author[label1]{Friedrich Wolf-Monheim} 

\affiliation[label1]{organization={Target Innovation AI},
            addressline={P.O. box 101611}, 
            city={Aachen},
            postcode={52064}, 
            state={NRW},
            country={Germany}}

\begin{abstract}
Next to decision tree and k-nearest neighbours algorithms deep convolutional neural networks (CNNs) are widely used to classify audio data in many domains like music, speech or environmental sounds.
To train a specific CNN various spectral and rhythm features like mel-scaled spectrograms, mel-frequency cepstral coefficients (MFCC), cyclic tempograms, short-time Fourier transform (STFT) chromagrams, constant-Q transform (CQT) chromagrams and chroma energy normalized statistics (CENS) chromagrams can be used as digital image input data for the neural network.
The performance of these spectral and rhythm features for audio category level as well as audio class level classification is investigated in detail with a deep CNN and the ESC-50 dataset with 2,000 labeled environmental audio recordings using an end-to-end deep learning pipeline.
The evaluated metrics accuracy, precision, recall and F\textsubscript{1}~score for multiclass classification clearly show that the mel-scaled spectrograms and the mel-frequency cepstral coefficients (MFCC) perform significantly better than the other spectral and rhythm features investigated in this research for audio classification tasks using deep CNNs.
\end{abstract}


\end{frontmatter}



\pagebreak[4]

\section{Introduction}
\label{sec1}

Audio classification and acoustic scene characterization algorithms are widely used in the area of machine learning today \cite{Piczak2015, Sainath2015, Barchiesi2015, Hershey2017, Dorfer2018}.
The diverse application areas for these algorithms are for example sound localization \cite{app11041519}, sound source separation \cite{app12020832} , speech recognition \cite{doi:10.1080/02564602.2015.1010611, Butko2011}, music classification \cite{NANNI201749}, differentiating between musical instruments \cite{Schubert2006DoesTB, app11146461}, audio scene characterization \cite{app12031569}, audio segmentation and sound event detection \cite{app12073293}, predictive maintenance \cite{app12094385}, surveillance \cite{Divakaran2005AudioAF} and bioacoustic monitoring \cite{Salamon2016TowardsTA}.
One common approach to classify audio signals with the aid of machine learning models like for example deep convolutional neural networks is by transforming the audio data in spectral and rhythm features like mel-scaled spectrograms, mel-frequency cepstral coefficients (MFCC), cyclic tempograms, short-time Fourier transform (STFT) chromagrams, constant-Q transform (CQT) chromagrams and chroma energy normalized statistics (CENS) chromagrams \cite{Gupta2017, Liu2010, Peeters2004, Sigtia2014}.
While these features have been extensively used in various audio classification tasks \cite{Lee2009, Panagakis2014, Stowell2013}, a systematic and comparative evaluation of their classification performance across different contexts is still lacking. Therefore this research aims to address the following key questions:

\begin{enumerate}[leftmargin=.6in]
\setlength{\itemsep}{5pt}

    \item How do different spectral and rhythm features influence the accuracy and robustness of audio classification models?
    
    \item Which feature yields the best classification performance for audio data?

    \item How do standard evaluation metrics such as accuracy, precision, recall and F\textsubscript{1} score vary depending on the choice of audio features?
    
\end{enumerate}

Comparing spectral and rhythm features in audio classification with deep CNNs is crucial for understanding which features best contribute to classification accuracy to finally optimize the model performance and to reduce computational complexity.
Different features capture distinct aspects of audio signals, which influences the ability of the deep CNN to learn patterns and generalize across tasks \cite{Huang2018, Sainath2015_2}.
This comparison also helps improve model robustness, interpretability and efficiency to ensure better performance in real-world applications.

Environmental sound classification (ESC) has been widely researched over the years, using both traditional machine learning and deep learning approaches \cite{Virtanen2018,Phan2019,Duan2019}. Classical methods have used techniques such as decision trees, support vector machines (SVMs) and k-nearest neighbors (k-NN) using hand-crafted features like Mel-frequency cepstral coefficients (MFCC) and spectral features \cite{rabiner1993fundamentals, bishop2006pattern, davis1980comparison, cortes1995support, altman1992introduction, fix1989discriminatory}.
However, deep learning models, particularly CNNs, have revolutionized this field by enabling end-to-end feature extraction and classification and recent advancements in ESC research have highlighted the effectiveness of CNNs in learning hierarchical representations from audio spectrograms, outperforming traditional approaches in terms of classification accuracy and robustness \cite{pons2017design, dai2017very, su2021esc, koizumi2020deep, gong2021ast}.
The ESC-50 dataset \cite{Piczak2015ESCDF} has developed into a benchmark dataset for environmental sound classification and widely adopted for evaluating machine learning models \cite{smith2018advancements, lee2019deep, garcia2020environmental, wang2021transformer, chen2022data, rodriguez2023ensemble, singh2024novel, zhang2025self}.
Comprising 2,000 labeled recordings across 50 diverse sound classes, the dataset is used for performance comparisons among various classification algorithms.
Due to its well-structured taxonomy and balanced class distribution, it has been extensively used for developing and testing CNN-based models, making it a good resource for audio classification research.
Compared to other datasets, ESC-50 maintains a standardized format, facilitating reproducible experiments and fair comparisons between different machine learning approaches.
Additionally, its broad adoption in the research community enables direct performance benchmarking with state-of-the-art models, making it a crucial resource for advancing environmental sound classification techniques \cite{mushtaq2020environmental, wang2024evaluation, jahangir2023deep}.

Deep CNNs have demonstrated exceptional performance in audio classification tasks by effectively capturing temporal and spectral features from input data \cite{Kumar2018, Liu2018, Choi2017}.
Various CNN architectures \cite{krizhevsky2012imagenet, he2016deep, huang2017densely}, including VGG-like networks \cite{simonyan2015very}, ResNet \cite{he2016deep} and hybrid models \cite{ashraf2023hybrid, ullo2020hybrid} incorporating recurrent layers have been explored for environmental sound classification tasks.
CNNs benefit from their ability to automatically learn hierarchical feature representations, thereby reducing the need for extensive manual feature engineering \cite{lee2016efficient}.
Feature selection plays a pivotal role in optimizing audio classification models, as different features capture distinct characteristics of sound signals \cite{Guyon2003}.
Spectral features such as mel-scaled spectrograms, short-time Fourier transform (STFT) spectrograms and constant-Q transform (CQT) representations capture frequency-domain information, while rhythm-based features like cyclic tempograms provide temporal structure insights \cite{Stevens1989, Brown1991, Schneider2002}.

The main objectives of this research are to systematically analyze and compare the classification performance of various spectral and rhythm features and to provide insights into their effectiveness for different audio classification tasks.
By conducting a rigorous evaluation using a machine learning model, this research tries to establish a quantitative understanding of the strengths and limitations of each feature type.
In this research the audio classification performance of the above-mentioned spectral and rhythm features is objectively analyzed in a systematic way using accuracy, precision, recall and F\textsubscript{1} score for multiclass classification.

The novelty of this work lies in its comprehensive and systematic comparison of spectral and rhythm features using multiple performance metrics across different classification tasks.
This research provides a holistic evaluation framework that can serve as a reference for future developments in audio classification.
The findings of this research contribute to optimizing feature selection for machine learning-based audio classification to finally improve the efficiency and accuracy of real-world applications.

\section{Methodology and Experiments}
\label{sec2}

\subsection{ESC-50 Dataset}
\label{subsec1}

For the experiments conducted within the scope of the research work the ESC-50 dataset \cite{Piczak2015ESCDF} was used.
The ESC-50 dataset is an audio data collection designed for environmental sound classification tasks \cite{guo2019audio, salamon2017deep}.
The dataset is an essential resource for researchers and developers working in the field of audio signal processing and machine learning \cite{herbordt2019comparison, dixon2006tut}.
The ESC-50 dataset is structured to support the development and evaluation of algorithms capable of recognizing environmental sounds and it consists of 2,000 labeled audio recordings.
These recordings are evenly distributed across 50 different classes with 40 samples each and each class is representing a unique environmental sound category.
These 50 different classes are grouped into five major categories. Therefore, each major audio category has 400 samples.
The categories range from natural sounds like rain, thunder and wind to human-made noises such as clock alarm, helicopter and chainsaw.
The 50 different classes are grouped in 5 different major categories (\Cref{tab:ESC50_classes}).

Each recording in the dataset has a duration of 5 seconds and is provided in a unified audio format (.wav-files) which ensures consistency and ease of use of the ESC-50 dataset.
The standardization with regards to audio format and the length of the individual audio data files enables straightforward preprocessing and analysis of the database.
The sounds included in the dataset are particularly from field recordings and sound effect libraries.
The ESC-50 dataset is licenced under a Creative Commons Attribution-NonCommercial 3.0 Unported (CC BY-NC 3.0) licence.
This means that researchers are free to share and adapt the dataset as long as they provide proper attribution and do not use it for commercial purposes.
The link to the GitHub repository of the ESC-50 dataset is given in the references section \cite{Piczak2015ESCDF}.
The ESC-50 dataset primarily contains sounds sampled at 44.1 kHz with 16-bit resolution, which is the standard for high-quality digital audio.
The sampling rate of 44.1 kHz ensures that frequencies up to 22.05 kHz are captured, covering most of the human hearing range, while the 16-bit resolution provides 65,536 discrete amplitude levels, offering a good signal-to-noise ratio (approximately 96 dB) and minimizing quantization noise.
Therefore, the quality level of the the audio recordings of the ESC-50 dataset is suitable for rigorous academic and professional applications.
The ESC-50 dataset is widely used in many different audio classification tasks, including but not limited to sound recognition \cite{guo2018audio}, environmental sound analysis \cite{salamon2017low} and machine learning model benchmarking \cite{guzhov2020sound} and it enables researchers to train, validate and test models designed for automatic sound recognition.
The ESC-50 dataset is a challenging benchmark for audio classification models to test their robustness and their ability to generalize to real-world audio data due to several factors.
It consists of 50 different classes covering a wide range of environmental sounds, such as animal noises, human actions, natural sounds and mechanical noises.
This diversity increases the complexity of distinguishing between classes.
Additionally, the dataset contains only 2,000 labeled audio clips (40 per class), which is relatively small for deep learning models and makes it prone to overfitting.
Each clip is only 5 seconds long, which requires models to effectively extract meaningful features from limited data.
Some sounds are very similar to each other, such as wind versus rain or laughing versus coughing, which makes audio classification more difficult.
The real-world audio recordings of the ESC-50 dataset often contain background noise and variations in recording conditions, such as differences in microphone quality and environmental acoustics, which challenge model generalization.
Furthermore, environmental sounds often have complex temporal structures and spectral characteristics, which makes feature extraction and classification more demanding.
Performance on this dataset is often used as a metric to gauge the effectiveness of new algorithms or approaches in the field of environmental audio classification.
The ESC-50 dataset is a valuable resource for the development of audio classification systems and provides a rich and varied collection of environmental sounds that can be used to challenge and refine algorithms and its structured format, wide range of sound categories and consistency makes it an ideal tool for cutting-edge research in audio signal processing \cite{guo2019audio, dai2019conv}.

\begin{table}
  \centering
  \begin{tabular}{P{4cm} P{4cm} P{4cm}}
  \toprule
  Animals (I)           & Natural soundscapes      & Human non-speech        \\
                        & and water sounds (II)   & sounds (III)             \\
  \midrule
  Dog (0)               & Rain (10)                & Crying baby (20)        \\
  Rooster (1)           & Sea waves (11)           & Sneezing (21)           \\
  Pig (2)               & Crackling fire (12)      & Clapping (22)           \\
  Cow (3)               & Crickets (13)            & Breathing (23)          \\
  Frog (4)              & Chirping birds (14)      & Coughing (24)           \\
  Cat (5)               & Water drops (15)         & Footsteps (25)          \\
  Hen (6)               & Wind (16)                & Laughing (26)           \\
  Insects (flying) (7)  & Pouring water (17)       & Brushing teeth (27)     \\
  Sheep (8)             & Toilet flush (18)        & Snoring (28)            \\
  Crow (9)              & Thunderstorm (19)        & Drinking, sipping (29)  \\  
  \bottomrule
  & & \\
  \end{tabular}

  \centering
  \begin{tabular}{P{6cm} P{6cm}}
  \toprule
  Interior/domestic sounds (IV) & Exterior/urban noises (V) \\
  \midrule
  Door knock (30)         & Helicopter (40)          \\
  Mouse click (31)        & Chainsaw (41)            \\
  Keyboard typing (32)    & Siren (42)               \\
  Door, wood creaks (33)  & Car horn (43)            \\
  Can opening (34)        & Engine (44)              \\
  Washing machine (35)    & Train (45)               \\
  Vacuum cleaner (36)     & Church bells (46)        \\
  Clock alarm (37)        & Airplane (47)            \\
  Clock tick (38)         & Fireworks (48)           \\
  Glass breaking (39)     & Hand saw (49)            \\  
  \bottomrule
  \end{tabular}
  \caption{Major audio categories I to V (400 samples each) as well as corresponding semantical classes 0~to~49 (40 samples each) of the ESC-50 dataset \cite{Piczak2015ESCDF} for environmental sound classification}
  \label{tab:ESC50_classes}  
\end{table}

In addition to the ESC-50 dataset, there are several other sound and environmental sound datasets available for research and machine learning tasks.
Notable ones include UrbanSound8K \cite{salamon2014dataset}, which contains 8,732 labeled sound clips from 10 different classes of urban sounds, commonly used for sound classification tasks.
Another large-scale dataset is AudioSet \cite{gemmeke2017audio}, with over 2 million human-labeled 10-second sound clips from YouTube, covering a wide range of categories like human sounds, animal sounds and environmental sounds.
VoxCeleb \cite{nagrani2017voxceleb, chung2018voxceleb2} is a dataset containing speech recordings from thousands of celebrities which is ideal for speaker recognition and verification tasks.
TUT Sound Events 2016 \cite{Mesaros2016_TUT_database} contains 25,000 audio samples designed for sound event detection and classification.
FSD50K \cite{Fonseca2022FSD50K} is a large dataset of 50,000 sound clips across 200 categories, which is useful for fine-grained sound classification tasks.
The DCASE (Detection and Classification of Acoustic Scenes and Events) \cite{Mesaros2018_DCASE} series offers datasets focused on sound event detection, classification and acoustic scene recognition.
The GTZAN Genre Collection \cite{tzanetakis2002musical, tzanetakis2002automatic} is a dataset of 1,000 audio clips (30 seconds long) labeled with 10 music genres, which is commonly used in music genre classification tasks.
Common Voice \cite{ardila2020common} is a large dataset of voice recordings from contributors worldwide, designed for training speech recognition systems.
ESC-10 \cite{Piczak2015ESCDF}, a smaller version of ESC-50 \cite{Piczak2015ESCDF}, contains 10 classes of environmental sounds.
LibriSpeech \cite{panayotov2015librispeech} is a large corpus of read English speech used for training and evaluating automatic speech recognition (ASR) systems.

\subsection{Experimental Setup}
\label{subsec2}

The experimental setup for this research follows a structured approach to ensure reproducibility and robust model performance.
The process consists of multiple steps which are used based on best practices in deep learning for audio classification \cite{abayomi2022data, li2020review, wei2020comparison}.
In a first step, the dataset is prepared.
The ESC-50 dataset (described in \Cref{subsec1}) consists of 2,000 environmental audio recordings in .wav format.
These files are split into a training set (1,600 audio files, 80~\% of the data) and a validation set (400 audio files, 20~\% of the data).
The split ensures that the model is evaluated on unseen data, which helps prevent overfitting and provides an unbiased estimate of its generalization capability.
The dataset split is performed in a stratified manner, preserving the distribution of classes across training and validation sets.
In a second step, feature extraction is performed to convert raw waveform data into meaningful representations.
Using the \textit{librosa} package \cite{McFee2015librosaAA} for audio and music signal analysis in \textit{Python}, each .wav-file is transformed into a set of spectral and rhythm features.
Spectral and rhythm features include mel-scaled spectrogram, mel-frequency cepstral coefficients (MFCC), cyclic tempogram, short-time Fourier transform (STFT) chromagram, constant-Q transform (CQT) chromagram and chroma energy normalized statistics (CENS) chromagram.
The decision to use these features is based on their effectiveness in capturing audio characteristics relevant to environmental sound classification.
The third step involves constructing a deep convolutional neural network (CNN) as described in more detail in \Cref{subsec4}.
The architecture is designed with multiple convolutional layers followed by pooling layers to extract hierarchical audio features.
The model also includes batch normalization to stabilize training and a dropout layer to mitigate overfitting.
The choice of the CNN architecture is based on its proven success in audio classification tasks, balancing complexity and efficiency.

For the training phase (fourth step), the \textit{Adam} optimizer \cite{Kingma2015} is selected due to its adaptive learning rate mechanism, which improves convergence speed and stability.
The loss function used is \textit{sparse categorical cross-entropy} \cite{Bishop:2006}, which is suitable for multiclass classification problems with integer-encoded labels.
The model is trained using a mini-batch size of 32 and an initial learning rate of 0.001, determined through empirical testing and inspired by \cite{medium_sound_classification} ensuring alignment with established methodologies in environmental sound classification research \cite{Munoz2018, Li2019, Zhang2020}.
To enhance model training and generalization, two callbacks are implemented:

\begin{enumerate}[leftmargin=.6in]
\setlength{\itemsep}{5pt}

    \item Learning Rate Reduction:
    The learning rate is reduced by a factor of 0.1 if the validation loss plateaus for two consecutive epochs.
    This adaptive adjustment helps the optimizer escape local minima and refine the model's performance.

    \item Early Stopping:
    Training is halted automatically, if the validation loss does not improve for six consecutive epochs, which prevents overfitting and reduces unnecessary computations.
    
\end{enumerate}

Only one set of experiments was performed because the chosen CNN model architecture was already optimized based on prior research and preliminary testing, as inspired and partly derived from \cite{medium_sound_classification}.
Given that CNNs have consistently demonstrated superior performance in audio classification tasks, conducting multiple experiments with alternative models was deemed unnecessary \cite{Pons2016, Lee2017}.
Instead, efforts were concentrated on refining hyperparameters and optimizing the training process to ensure stable performance on the ESC-50 dataset.
The experimental setup includes only a training and a validation set to streamline the experimental process and the validation set is used for model optimization, including hyperparameter tuning and early stopping.
Therefore, the dataset is split into training (80~\%) and validation (20~\%) sets, with no separate test set, and the validation serves as a proxy for testing.
In \Cref{tab:training_metrics} the training metrics for the various spectral and rhythm features are shown.

\begin{table}
  \centering
  \begin{tabular}{P{2.5cm} P{2.1cm} P{2.1cm} P{2.1cm} P{2.1cm}}
  \toprule
  Spectral/ & Loss & Accuracy & Validation & Validation \\
  rhythm    &      &          & loss       & accuracy   \\
  feature   & [-]  & [~\%]    & [-]        & [~\%]   \\
  \midrule
  Mel-scaled spectrogram & 0.204 & 94.7 & 1.869 & 61.8 \\
  \midrule
  MFCC & 0.213 & 93.9 & 2.180 & 58.8 \\
  \midrule
  Cyclic tempogram & 2.040 & 41.5 & 3.040 & 23.3 \\
  \midrule
  STFT chromagram & 0.450 & 85.6 & 4.833 & 21.5 \\
  \midrule
  CQT chromagram & 0.726 & 77.7 & 3.949 & 21.3 \\
  \midrule
  CENS chromagram & 1.535 & 54.8 & 3.953 & 14.0 \\
  \bottomrule
  \end{tabular}
  \caption{Training metrics for various spectral and rhythm features}
  \label{tab:training_metrics}  
\end{table}

\subsection{Spectral and Rhythm Features}
\label{subsec3}

Six spectral and rhythm features are investigated in this research in terms of the audio classification performance using deep convolutional neural networks (CNNs).

\begin{figure}
\includegraphics[width=\textwidth]{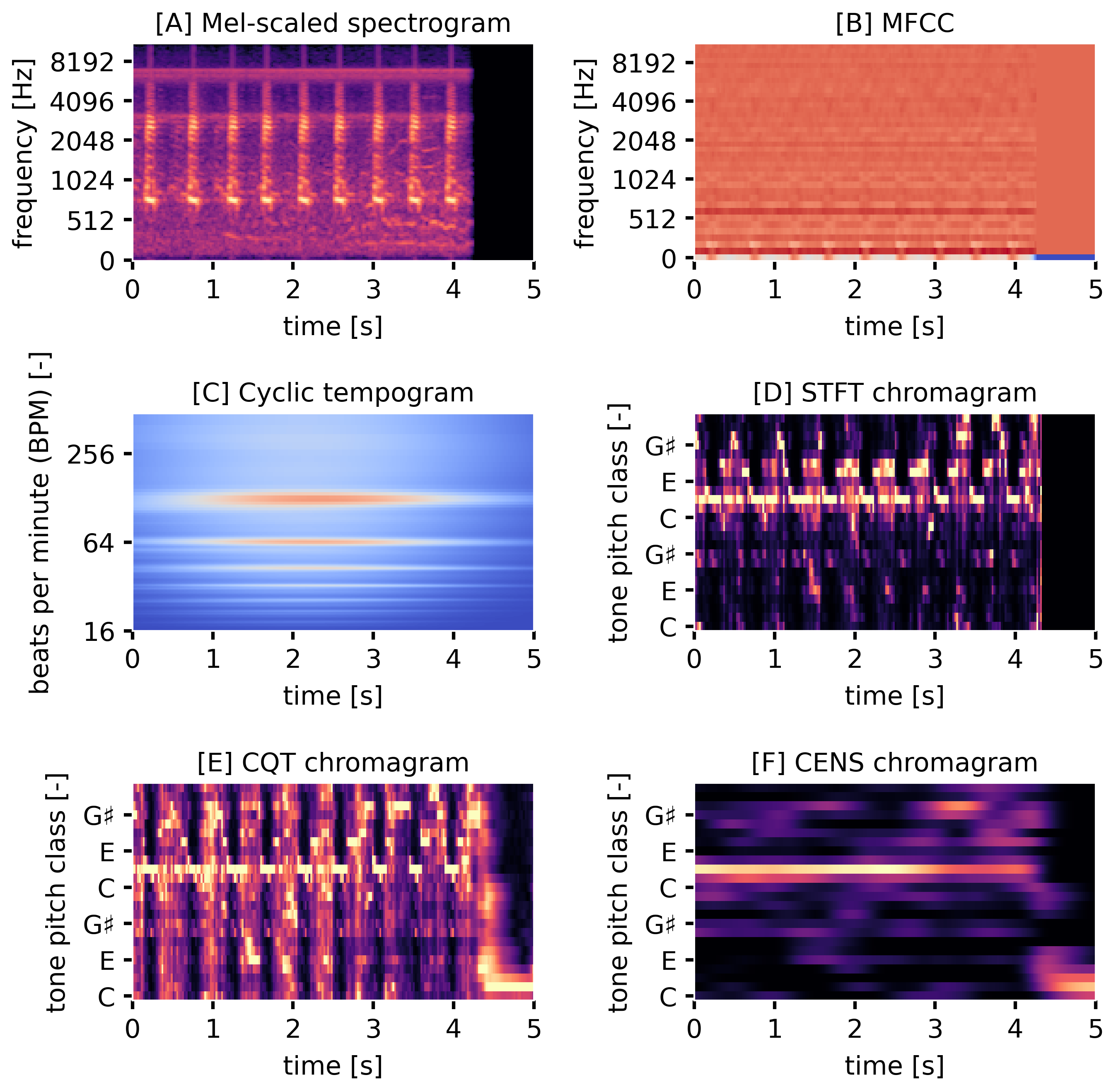}
\caption{Spectral and rhythm features: [A] Mel-scaled spectrogram, [B] Mel-frequency cepstral coefficients (MFCC), [C] Cyclic tempogram, [D] Short-time Fourier transform (STFT) chromagram, [E] Constant-Q transform (CQT) chromagram and [F] Chroma energy normalized statistics (CENS) chromagram; audio clip example form ESC-50 dataset (class: 'frog', 4)}
\label{fig:6_features}
\end{figure}

The first feature is the mel-scaled spectrogram (\Cref{fig:6_features} [A]) \cite{jang2019music,Waldekar2020-ec,Shen2018-kr}  as a special version of a spectrogram \cite{Jurafsky2009,Yu2014AutomaticSR,6857341}.
Spectrograms can be generated from sound signals using Fourier transforms which decompose the audio signals into their constituent frequencies.
They display the amplitude of each frequency (y-axis) present in the audio signal over time (x-axis).
Different colors are used to indicate the amplitude of each frequency.
The brighter the color of the plot the higher the energy of the audio signal.
In a mel-scaled spectrogram the frequencies on the y-axis are converted to the mel scale.
The advantages of mel-scaled spectrograms are that they capture perceptually relevant frequency information, are useful for speech and music analysis (e.g. genre classification or speaker recognition) and robust to noise compared to raw spectrograms.
The disadvantages are that they are high-dimensional, requiring more computing capacity, lose phase information, which can be important for some applications, and are not directly suited for pitch or harmonic content analyses \cite{Oppenheim1999}.

A mel-scaled spectrogram is obtained by mapping a spectrogram onto the mel scale using a filter bank. First the Short-Time Fourier Transform (STFT) is calculated:

\begin{equation}
    X(m,k) = \sum_{n=0}^{N-1} x(n) w(n - mH) e^{-j 2\pi k n / N}
\end{equation}

where:

\begin{itemize}
    \item $X(m,k)$ is the STFT at frame $m$ and frequency bin $k$,
    \item $x(n)$ is the input signal,
    \item $w(n)$ is the analysis window and
    \item $H$ is the hop size.
\end{itemize}

In a second step the power spectrogram is computed:

\begin{equation}
    P(m,k) = |X(m,k)|^2
\end{equation}

Finally, the mel filter bank is applied:

\begin{equation}
    M(m,j) = \sum_{k} H_j(k) P(m,k)
\end{equation}

where $H_j(k)$ represents the mel filter bank \cite{davis1980comparison2}.

The second feature are mel-frequency cepstral coefficients (MFCC) (\Cref{fig:6_features} [B]) \cite{Logan2000MelFC, Xu2004HMMBasedAK, SAHIDULLAH2012543, Abdulsatar_2019, Zheng2001}.
MFCCs are calculated by a sequence of steps.
In a first step the input signal is divided into blocks or windows (e.g. Hamming window function to avoid edge effects).
Overlapping windows are common.
The second step is a (discrete) Fourier transformation of each individual window.
This transforms the convolution of the excitation signal and the impulse response into a multiplication.
In a third step the magnitude spectrum is generated.
The fourth step is the logarithmization of the magnitude spectrum.
This transforms the multiplication of the excitation signal and the impulse response into an addition.
Subsequently, the number of frequency bands (e.g. 256) are reduced by merging (e.g. to 40) in a fifth step, which is a mapping to the mel scale in discrete steps using triangular filters (effectively bandpass filtering).
Finally, in a sixth step a decorrelation by either a discrete cosine transform or a principal component analysis (also called Karhunen-Loève transform) is conducted.
The advantages of mel-frequency cepstral coefficients (MFCC) is that they are effective in speech recognition and music genre classification, offering a compact representation with only a few coefficients needed while helping to distinguish timbral characteristics.
However, they have disadvantages like that they are sensitive to noise and reverberation, lose detailed harmonic and pitch information and require proper parameter tuning for optimal performance.
MFCC is computed from the mel-scaled spectrogram by applying a logarithm and a discrete cosine transform (DCT). In a first step the logarithmic mel-scaled spectrogram is computed:

\begin{equation}
S(m,j) = \log M(m,j)
\end{equation}

Subsequently, the DCT is applied to decorrelate features:

\begin{equation}
C(n) = \sum_{j=0}^{J-1} S(m,j) \cos \left[ \frac{\pi n (j + 0.5)}{J} \right]
\end{equation}

where \( C(n) \) are the MFCC coefficients \cite{Logan2000MelFC}.

Similar to spectrograms, tempograms are time-tempo representations of time-dependent audio signals.
A tempogram encodes for instance the tempo of a music audio signal over time.
To generate a tempogram, an audio signal is subdivided into time intervals and the tempo, the pulse respectively the rhythmic information is analyzed.
Depending on the specific analysis of an audio signal, the tempo can be given in beats per minute (BPM) or any other rhythmic unit.
Graphical representations of tempograms typically show time (e.g. in seconds) on the x-axis and tempo (e.g. in BPM) on the y-axis.
The cyclic tempograms (\Cref{fig:6_features} [C]) used in this research are generated according to \cite{5495219}.
The advantages of cyclic tempograms are that they are good for analyzing tempo-related musical features, robust to variations in absolute tempo and useful for genre classification and beat tracking.
The disadvantages are that they are not effective for signals without clear rhythmic structure, have high computational cost and may lose harmonic and timbral details.

A cyclic tempogram represents the periodicity of a rhythm with respect to beats.
To compute a cyclic tempogram \cite{Grosche2010} first the autocorrelation of onset strength is calculated:

\begin{equation}
R(\tau) = \sum_{m} O(m) O(m+\tau)
\end{equation}

where \( O(m) \) is the onset envelope and \( \tau \) is the lag.

Then the cyclic representation is mapped using the Fourier transform:

\begin{equation}
T_c(f) = \sum_{\tau} R(\tau) e^{-j 2 \pi f \tau}
\end{equation}

A chromagram is a graphical visualization of a time-dependent audio signal to extract individual tone pitches over time.
In contrast to audio spectrograms like mel-scaled spectrograms showing the energy of audio signals over frequency and time, audio chromagrams specifically focus in the tonal content of audio signals.
To derive a chromagram from a specific audio signal three individual steps are necessary.
In a first step the individual tone pitches are calculated.
To do so the audio signal is cut into individual time frames with typical lengths of 20 to 100 milliseconds.
By using appropriate methods like for example Fourier transform or special tone pitch estimation algorithms such as auto-correlation the tone pitches can be calculated for all time frames.
In a second step the tone pitch visualization is prepared.
A vector representing the tone pitch distribution in the audio signal is generated.
Typically, the vector contains values for any tone pitch or semitone step in the musical spectrum.
The individual values of the vector can either be binary or continuous.
Finally, the graphical representation of the chromagram is generated in a third step by aggregating all vectors of the source audio signal over time.
The first axis of a chromagram represents time and the second axis of a chromagram represents the individual tone pitches.
The brightness or color in a specific area of a chromagram is a measure of the strength or frequency of a respective tone pitch in a specific time frame \cite{Fujishima1999, Bello2005}.

Short-time Fourier transform (STFT) chromagrams are a specific type of chromagram leveraging the advantages of short-time Fourier transform algorithms for audio signal processing to extract the frequency information of an audio signal over time.
The short-time Fourier transform decomposes the source audio signal into its spectral components for each time frame to finally analyze the frequency distribution of the audio signal over time.
The tone pitches for the individual time frames are determined using tone pitch estimation algorithms.
To generate a graphical representation of an STFT chromagram
the distributed tone pitches are plotted over time and frequency.
Typically, the x-axis of an STFT chromagram represents time, the y-axis represents the tone pitches and the color or brightness of the chromagram represents the intensity and/or frequency of the respective tone pitches.
The STFT chromagrams (\Cref{fig:6_features} [D]) used in this research are generated according to \cite{chromagram}.
The advantages of STFT chromagrams are that they are good for analyzing harmonic and tonal content, effective for chord recognition and key detection and preserve temporal resolution better than CQT chromagrams.
The disadvantages are that they are sensitive to tuning deviations and noise, the frequency resolution is limited at low frequencies and that they require post-processing for applications like key detection.

STFT chromagrams are computed by mapping spectral energy to pitch classes. In a first step the STFT power spectrogram is calculated:

\begin{equation}
P(m, k)
\end{equation}

Subsequently, the frequencies are mapped to chroma bins:

\begin{equation}
C(m, p) = \sum_{k \in K(p)} W(p, k) P(m, k)
\end{equation}

where \( W(p, k) \) is a weight matrix for chroma mapping \cite{Mller2011ChromaTM, Oppenheim1999}.

Constant-Q transform (CQT) chromagrams (\Cref{fig:6_features} [E]) are another spectral feature.
CQT chromagrams leverage the advantages of the constant-Q transform algorithms to generate chromagrams.
Constant-Q transform is related to the Fourier transform and very closely related to the complex Morlet wavelet transform \cite{Brown1991CalculationOA, Cwitkowitz2019EndtoEndMT, Brown1992AnEA}.
The constant-Q transform is characterized by the fact that the bandwidth and the sampling density can differ from each other relative to the frequency. 
The individual time frames are constructed and applied directly in the frequency domain.
Distinct time frames exhibit different center frequencies and bandwidths.
However, the ratio between center frequency and bandwidth remains constant.
Maintaining a constant ratio between center frequency and bandwidth presupposes that the time resolution improves at higher frequencies and the frequency resolution improves at lower frequencies.
Due to the blurring principle the time delays for each time frame depend on the bandwidth.
The advantages of CQT chromagrams are that they have a higher frequency resolution at lower pitches, are useful for harmonic analysis and key detection and better suited for music applications than STFT chromagrams.
The disadvantages are that they have higher computational cost than STFT chromagrams, lose some temporal resolution due to longer window sizes and can be affected by tuning variations.
CQT chromagrams use a logarithmic frequency resolution \cite{Brown1991CalculationOA}. First the constant-Q transform (CQT) is conducted:

\begin{equation}
X_{\text{CQ}}(m,k) = \sum_{n=0}^{N-1} x(n) g_k(n) e^{-j 2 \pi k n / N}
\end{equation}

where \( g_k(n) \) are CQT basis filters.

In a second step the chroma mapping is computed:

\begin{equation}
C(m,p) = \sum_{k \in K(p)} W(p,k) \left| X_{\text{CQ}}(m,k) \right|
\end{equation}

A third variant of chromagrams are chroma energy normalized statistics (CENS) chromagrams.
CENS chromagrams have the advantage that they offer a robust and scalable representation of tonal structures in time-dependent audio signals.
The basis of CENS chromagrams is the chroma energy normalization (CEN).
CEN normalizes the energy values of the chromagrams to increase the robustness of a specific audio signal against variations of sound volume and tone color.
CENS chromagrams can be used for tone pitch analysis and audio signal information extraction since they do not only extract tone pitches but are also insensitive against variations of tone pitches and sound dynamics.   
To generate CENS chromagrams two additional steps have to be taken after computing the constant-Q transform as described above.
In a first additional step an L1 normalization is computed of each individual chromagram vector.
In a second additional step a quantization step of the amplitudes is conducted based on 'log-like' amplitude thresholds.
The CENS chromagrams (\Cref{fig:6_features} [F]) used in this research are generated according to \cite{Mller2011ChromaTM}.
The advantages of CENS chromagrams are that they are robust to dynamics and noise variations, useful for large-scale music structure analysis and retrieval and help in key and chord recognition tasks.
The disadvantages are the loss of fine spectral details, the drawbacks for applications requiring precise pitch tracking and that the smoothing may reduce discriminative power in some cases.
CENS chromagrams are derived from CQT chromagrams and further smoothed \cite{Muller2005, Mller2011ChromaTM}. In a first step the CQT Chromagram $C(m,p)$ is computed.
In a second step energy normalization is applied:

\begin{equation}
    \hat{C}(m,p) = \frac{C(m,p)}{\sum_{p} C(m,p)}
\end{equation}

Finally, smoothing (e.g. mean filter) is applied to reduce noise and fluctuations in the chromagram data by averaging the values within a local window.
This is typically done to enhance the clarity of the CENS chromagram representation by mitigating high-frequency variations or sudden changes in the chroma features that could be the result of noise or minor inconsistencies in the data.

In \Cref{tab:feature_comparison} the various spectral and rhythm features are summarized and compared with each other.

\begin{table}
  \centering
  \begin{tabular}{P{2.5cm} P{3.15cm} P{3.15cm} P{3.15cm}}
  \toprule
  Feature & Best for & Advantages & Disadvantages \\
  \midrule
  Mel-scaled spectrogram & Speech and music analysis & Perceptually relevant, noise robust & High-dimensional, phase loss \\
  \midrule
  MFCC & Speech and genre classification & Compact, good timbral representation & Noise-sensitive, loses harmonic details \\
  \midrule
  Cyclic tempogram & Rhythm and tempo analysis & Tempo robustness, useful for beat tracking & Computationally expensive, not for non-rhythmic signals \\
  \midrule
  STFT chromagram & Chord and key detection & Good harmonic representation, keeps temporal details & Sensitive to tuning, low frequency resolution \\
  \midrule
  CQT chromagram & Harmonic analysis & Better low-frequency resolution & Computationally expensive, lower temporal resolution \\
  \midrule
  CENS chromagram & Large-scale music analysis & Noise robust, good for key detection & Loses fine spectral details \\
  \bottomrule
  \end{tabular}
  \caption{Summary of spectral and rhythm features with advantages and disadvantages as well as main target application area}
  \label{tab:feature_comparison}  
\end{table}

\subsection{Deep Convolutional Neural Network (CNN)}
\label{subsec4}

Deep Convolutional Neural Networks (CNNs) \cite{krizhevsky2012imagenet, lecun1998gradient, simonyan2015very} are a class of deep learning models primarily used for processing matrix-like data such as digital images and spectrograms.
They are very useful for tasks such as image classification, object detection and speech recognition.
Unlike fully connected neural networks, CNNs are designed to automatically detect spatial hierarchies in the input data using specialized layers like convolutional layers, pooling layers and normalization layers.
They are built to recognize patterns in a localized manner, which makes them particularly efficient for tasks where patterns are important but spatial relationships also need to be preserved.

CNNs typically consist of three main types of layers:

\begin{enumerate}[leftmargin=.6in]
\setlength{\itemsep}{5pt}

    \item Convolutional layers:
    These layers apply a set of filters (kernels) to the input data, enabling the network to learn spatial hierarchies and detect features such as edges, textures and shapes.
    
    \item Pooling layers:
    These layers reduce the spatial dimensions of the input, summarizing the important features while reducing computational complexity and overfitting.

    \item Fully connected layers:
    These layers process the high-level features extracted by the convolutional and pooling layers and make the final predictions.
    
\end{enumerate}

The deep CNN architecture used in this research is designed to process spectral and rhythm features, which are two-dimensional representations of audio signals over time.
The network architecture includes several layers that work together to extract meaningful features from the spectral and rhythm features and ultimately make a classification prediction.
The deep CNN has the following layers:

\begin{enumerate}[leftmargin=.6in]
\setlength{\itemsep}{5pt}

    \item Batch normalization layer:
    This layer is used to normalize the input data along the frequency axis.
    It helps in stabilizing the training process by ensuring that the input values have a consistent mean and variance.
    This improves the convergence and generalization of the model.
    
    \item 2D convolution layer (filter: 64, height: 3 pixels, width: 3 pixels, activation function: \textit{rectified linear unit (ReLU)}, padding: \textit{same}):
    This convolutional layer applies 64 filters of size 3x3 to the input data.
    The filters learn spatial features such as edges or textures.
    The \textit{ReLU} activation function used introduces non-linearity and helps the network learn more complex patterns.
    Padding is set to \textit{same} to ensure that the spatial dimensions of the output are the same as the input.
    
    \item Maximum pooling layer for 2D spatial data (pooling window: 2x2):
    This layer performs maximum pooling with a window of size 2x2.
    Pooling reduces the spatial size of the data by selecting the maximum value from each region of the input, which helps to decrease the computational cost and make the network invariant to small translations of features.
    
    \item 2D convolution layer (filter: 128, height: 3 pixels, width: 3 pixels, activation function: \textit{rectified linear unit}, padding: \textit{same}):
    The second convolutional layer applies 128 filters of size 3x3 to the output from the previous pooling layer.
    Again, the \textit{ReLU} activation function is used and padding is set to \textit{same}.
    This layer extracts higher-level features from the input data.
    
    \item Maximum pooling layer for 2D spatial data (pooling window: 2x2):
    A second pooling layer is applied with a 2x2 window to further downsample the spatial dimensions of the feature maps.

    \item 2D convolution layer (filter: 256, height: 3 pixels, width: 3 pixels, activation function: \textit{rectified linear unit}, padding: \textit{same}):
    This layer applies 256 filters of size 3x3, again using the \textit{ReLU} activation function and \textit{same} padding.
    It extracts even more complex and abstract features from the input data.

    \item Maximum pooling layer for 2D spatial data (pooling window: 2x2):
    Another pooling layer reduces the spatial dimensions and focuses on the most important features learned by the previous convolutional layers.

    \item 2D convolution layer (filter: 256, height: 3 pixels, width: 3 pixels, activation function: \textit{rectified linear unit}, padding: \textit{same}):
    This layer also uses 256 filters of size 3x3 and applies \textit{ReLU} activation and \textit{same} padding.
    It continues the process of extracting high-level features from the input data.

    \item Maximum pooling layer for 2D spatial data (pooling window: 2x2):
    Another pooling layer is applied with a 2x2 window to further compress the feature maps.

    \item Flatten layer:
    This layer converts the 2D feature maps into a 1D vector to make the data suitable for input into a fully connected layer.
    It essentially unrolls the pooled feature maps into a single long 1D vector.

    \item Regular densly-connected neural network layer (units: 256, activation function: \textit{rectified linear unit}):
    This dense layer has 256 units and uses \textit{ReLU} activation.
    It processes the high-level features extracted from the previous layers and learns complex patterns that are relevant for the final classification.

    \item Dropout layer with a rate of 0.5:
    This layer is included to prevent overfitting during training.
    It randomly disables 50~\% of the neurons during each training iteration to force the network to generalize better and not rely on specific neurons too much.

    \item Regular densly-connected neural network layer (units: 50, activation function: \textit{softmax}):
    The final fully connected layer converts the 1D output vector of values to a probability distribution.
    The \textit{softmax} function produces a probability distribution over the possible classes.
    This allows the deep CNN to make a probabilistic prediction for each input dataset.
    
\end{enumerate}

In summary the deep convolutional neural network (CNN) used in this research is optimized to classify spectral and rhythm features efficiently.
It begins with a batch normalization layer to normalize the input data to finally stabilize the training process.
The multiple 2D convolution layers with increasing filter sizes (64, 128 and 256) in combination with the maximum pooling layers are used to extract increasingly complex features from the input data.
The convolution layers utilize \textit{ReLU} activation functions to prevent vanishing gradients and to promote faster convergence.
The pooling layers reduce the spatial dimensions of the feature maps, decrease the computational complexity and improve generalization by focusing on the most important features.
The flatten layer converts the input data into a 1D vector suitable for a fully connected layer, which further abstract the features for classification.
To prevent overfitting, dropout is applied with a rate of 0.5, ensuring the model generalizes well to new data.
The final softmax layer converts the output into probabilities, corresponding to the 50 classification categories.
The hyperparameters, such as the number of filters and the dropout rate, are optimized through experimentation to balance model complexity and computational efficiency for accurate and robust classification of spectral and rhythm features.

\subsection{Target Metrics to Describe Audio Classification Performance}
\label{subsec5}

To compare the six spectral and rhythm features described in \Cref{subsec3} in terms of the audio classification performance using the deep convolutional neural network described in \Cref{subsec4} accuracies, precisions, recall values and F\textsubscript{1}~scores \cite{Powers2011, Kuhn2013, Sokolova2009} for the classification performance on audio category and audio class levels are calculated.

Accuracy measures the fraction of correctly predicted class labels out of the total number of samples.
It is defined as:

\begin{equation}
\text{Accuracy} = \frac{\text{Number of Correct Predictions}}{\text{Total Number of Samples}}
\end{equation}

True Positives (TP) and True Negatives (TN) are correctly classified samples.
False Positives (FP) and False Negatives (FN) are incorrectly classified samples.
The accuracies are calculated according to \Cref{eqn:accuracy} as fractions of correct class predictions over $n_{samples}$ where $1(x)$ is the indicator function.
If the entire set of predicted class labels strictly matches with the true set of class labels, then the accuracy is 100~\%, otherwise it is 0~\%.
$\hat{y}_i$ is the predicted class label of the $i$-th audio sample and $y_i$ is the corresponding ground truth class label \cite{Hastie2009}.

\begin{equation}
\label{eqn:accuracy}
Accuracy(y,\hat{y})=\frac{1}{n_{samples}} \sum_{i=0}^{n_{samples}-1} 1(\hat{y}_i=y_i)
\end{equation}

Precision indicates how many of the predicted positive samples are actually correct:

\begin{equation}
\text{Precision} = \frac{TP}{TP + FP}
\end{equation}

The precisions are calculated by \Cref{eqn:precision} and the average precisions are calculated according to \Cref{eqn:precision_average}.
$y$ is the set of true (audio sample, audio label) pairs, $\hat{y}$ is the set of predicted (audio sample, audio label) pairs, $L$ is the set of audio labels and $y_l$ is the subset of $y$ with the audio label~$l$.
$P(A,B)$ is given by \Cref{eqn:precision_aux}.
In \Cref{eqn:precision}, \Cref{eqn:precision_average} and \Cref{eqn:precision_aux} $P(A,B):=0$ for $B = \emptyset$ \cite{Kubat2015, Manning2008}.

\begin{equation}
\label{eqn:precision}
\langle P(y_l,\hat{y}_l) | l \in L \rangle
\end{equation}

\begin{equation}
\label{eqn:precision_average}
\frac{1}{|L|} \sum\nolimits_{l\in L} P(y_l,\hat{y}_l)
\end{equation}

\begin{equation}
\label{eqn:precision_aux}
P(A,B):= \frac{|A \cap B|}{|B|} \ for \ some \ sets \ A \ and \ B
\end{equation}

Recall (or sensitivity) measures how many actual positive samples were correctly identified.
It is given by:

\begin{equation}
\text{Recall} = \frac{TP}{TP + FN}
\end{equation}

The recall values are calculated by \Cref{eqn:recall} and the average recall values are calculated according to \Cref{eqn:recall_average}.
$R(A,B)$ is given by \Cref{eqn:recall_aux}.
In \Cref{eqn:recall}, \Cref{eqn:recall_average} and \Cref{eqn:recall_aux} $R(A,B):=0$ for $A = \emptyset$ \cite{Manning2008}.

\begin{equation}
\label{eqn:recall}
\langle R(y_l,\hat{y}_l) | l \in L \rangle
\end{equation}

\begin{equation}
\label{eqn:recall_average}
\frac{1}{|L|} \sum\nolimits_{l\in L} R(y_l,\hat{y}_l)
\end{equation}

\begin{equation}
\label{eqn:recall_aux}
R(A,B):= \frac{|A \cap B|}{|A|} \ for \ some \ sets \ A \ and \ B
\end{equation}

F\textsubscript{1} score is the harmonic mean of precision and recall, balancing both metrics.
It is given by:

\begin{equation}
F1 = \frac{2 \times \text{Precision} \times \text{Recall}}{\text{Precision} + \text{Recall}}
\end{equation}

A high F\textsubscript{1} score indicates a good balance between precision and recall, reducing both FP and FN.

The F\textsubscript{1} scores are calculated by \Cref{eqn:f1_score} and the average F\textsubscript{1}~scores are calculated according to \Cref{eqn:f1_score_average}.
$F_1(A,B)$ is given by \Cref{eqn:f1_score_aux} \cite{van1979information, Powers2011}.

\begin{equation}
\label{eqn:f1_score}
\langle F_1(y_l,\hat{y}_l) | l \in L \rangle
\end{equation}

\begin{equation}
\label{eqn:f1_score_average}
\frac{1}{|L|} \sum\nolimits_{l\in L} F_1(y_l,\hat{y}_l)
\end{equation}

\begin{equation}
\label{eqn:f1_score_aux}
F_1(A,B):= 2 \cdot \frac{P(A,B) \times R(A,B)}{P(A,B) + R(A,B)}
\end{equation}

\section{Experimental Results}
\label{sec3}

\begin{figure}
\includegraphics[width=\textwidth]{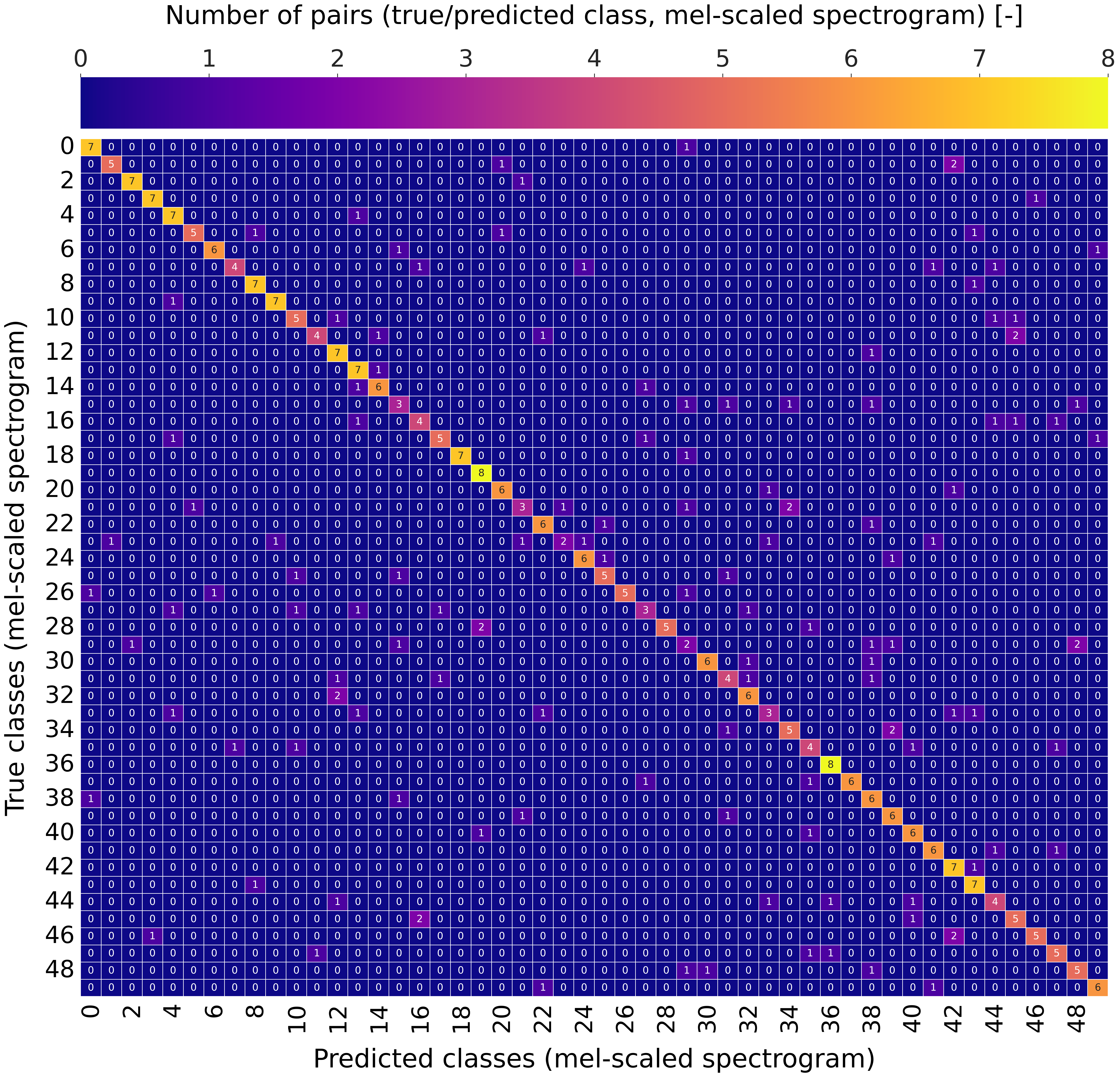}
\caption{Confusion matrix for ground truth audio class labels versus predicted audio class labels using mel-scaled spectrograms as spectral features}
\label{fig:confusion_matrix_melspec}
\end{figure}

To get an overview of the audio classification results on a class level, \Cref{fig:confusion_matrix_melspec} shows the confusion matrix for ground truth audio class labels versus predicted audio class labels using mel-scaled spectrogram features as an example.
Since 8 audio clips per class where sampled randomly out of the 2,000 labeled audio recordings of the ESC-50 dataset to generate the validation set of 400 audio recording (50 classes with 8 audio clips each) in total, the maximum number of pairs (true/predicted class) per box in the confusion matrix is 8.
The good performance of the mel-scaled spectrogram feature for the audio classification task can be seen by the high numbers on the main diagonal of the confusion matrix.

\begin{figure}
\includegraphics[width=\textwidth]{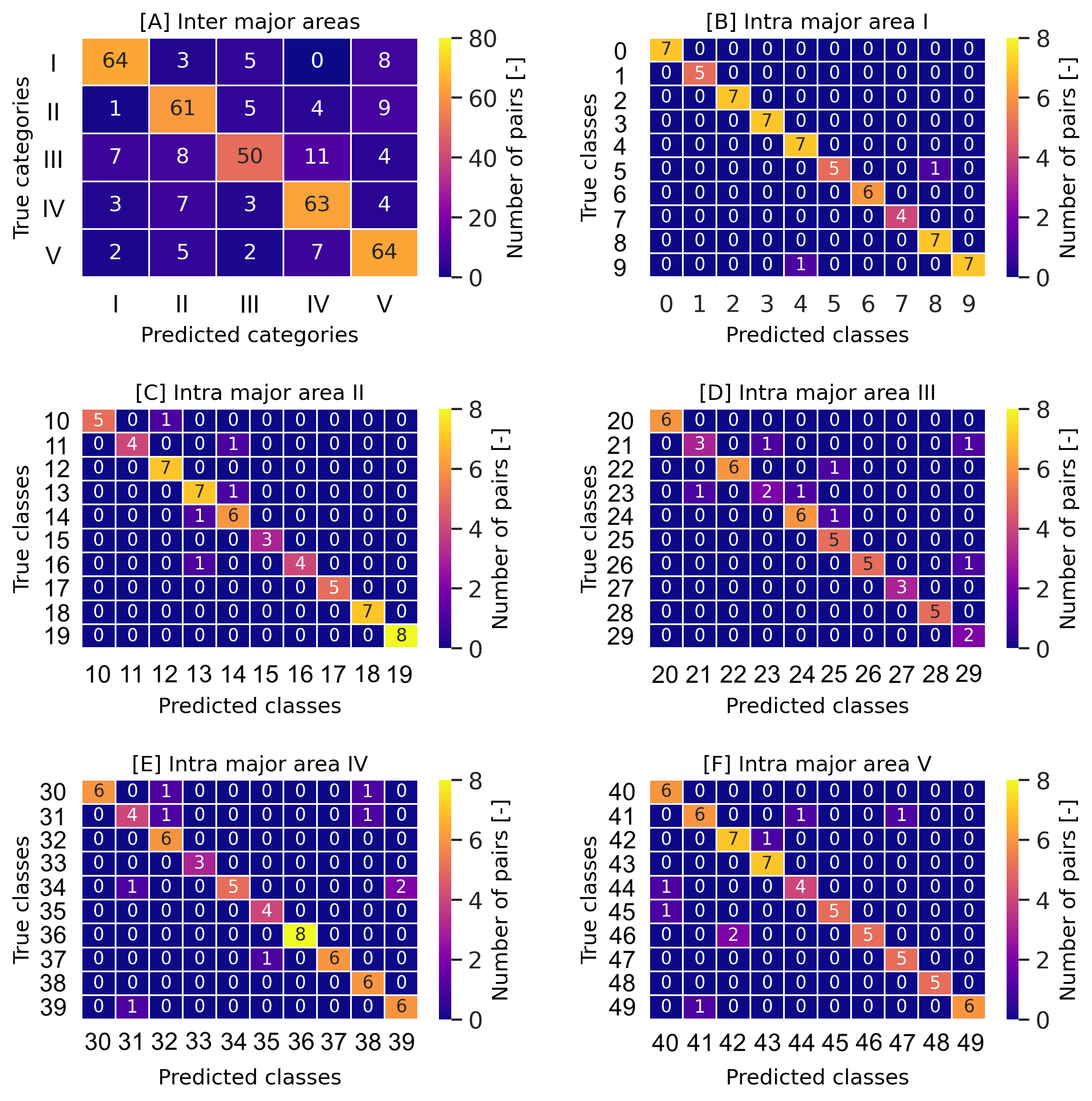}
\caption{Confusion matrices on audio category level ([A], inter major areas) and on audio class level ([B]~-~[F], intra major areas I~-~V) using mel-scaled spectrograms as spectral features}
\label{fig:cm_melspec_2}
\end{figure}

In \Cref{fig:cm_melspec_2} the confusion matrices on audio category level [A] as well as on audio class level for the individual main audio categories ([B] category I: animals, classes 0~-~9, [C] category II: natural soundscapes and water sounds, classes 10~-~19, [D] category III: human non-speech sounds, classes 20~-~29, [E] category IV: interior/domestic sounds, classes 30~-~39, [F] category V: exterior/urban noises, classes 40~-~49) are shown.

\begin{figure}
\includegraphics[width=\textwidth]{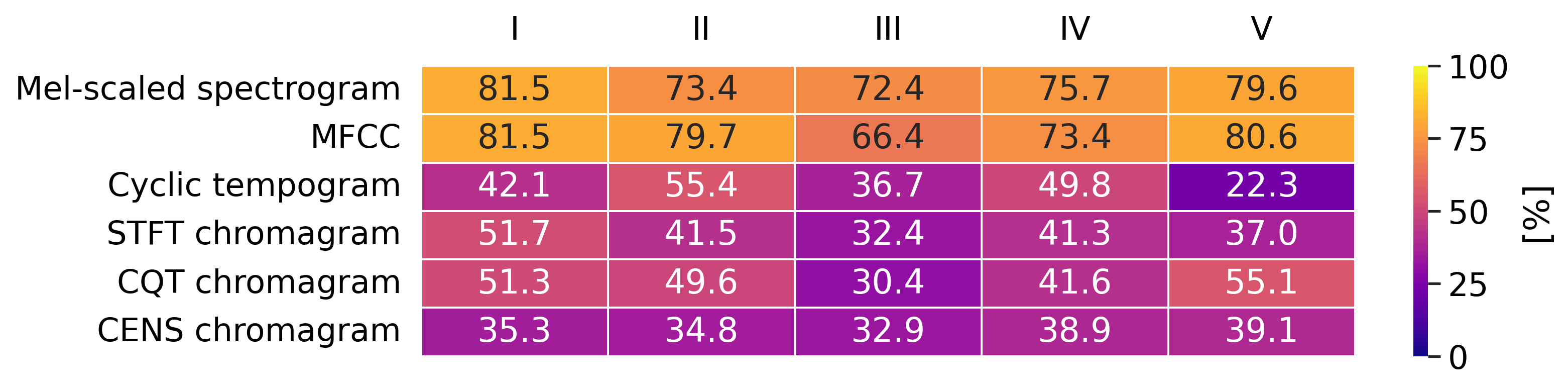}
\caption{Heatmap for accuracies in [\%] on audio category level (I~-~V) for six spectral and rhythm features}
\label{fig:heat_accuracy_cat_A}
\end{figure}

\begin{figure}
\includegraphics[width=\textwidth]{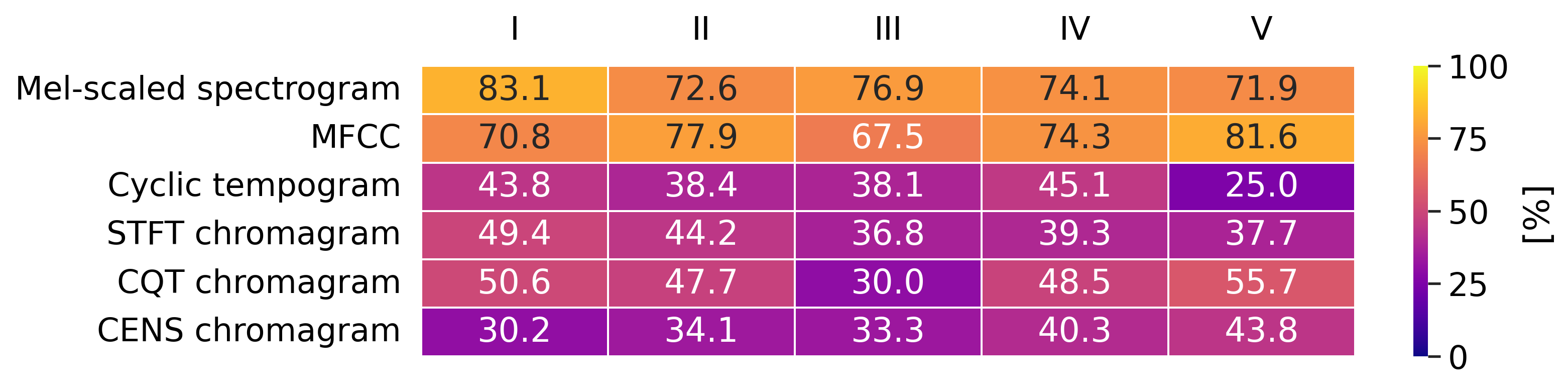}
\caption{Heatmap for precisions in [\%] on audio category level (I~-~V) for six spectral and rhythm features}
\label{fig:heat_precision_cat_A}
\end{figure}

\begin{figure}
\includegraphics[width=\textwidth]{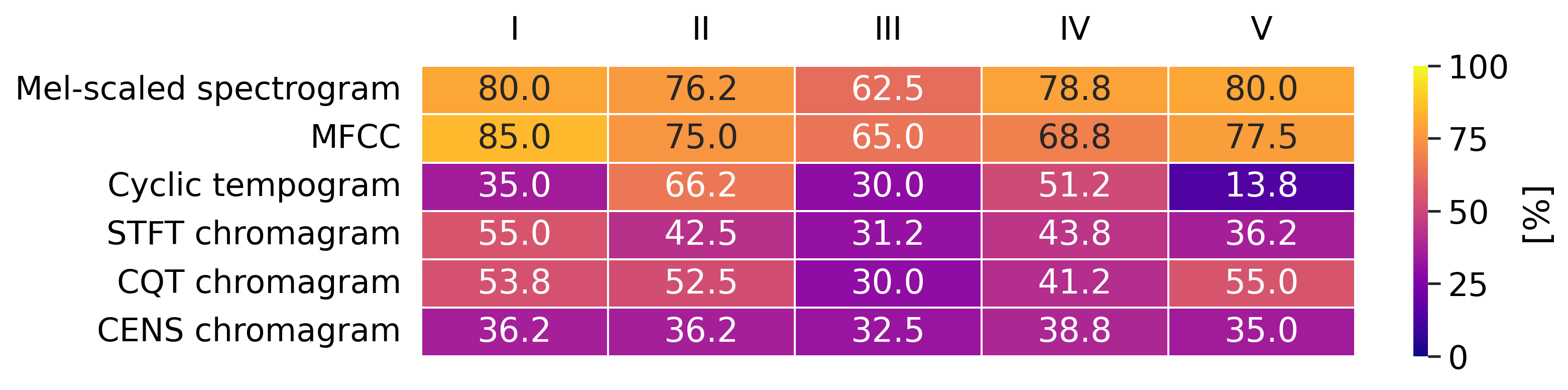}
\caption{Heatmap for recall values in [\%] on audio category level (I~-~V) for six spectral and rhythm features}
\label{fig:heat_recall_cat_A}
\end{figure}

\begin{figure}
\includegraphics[width=\textwidth]{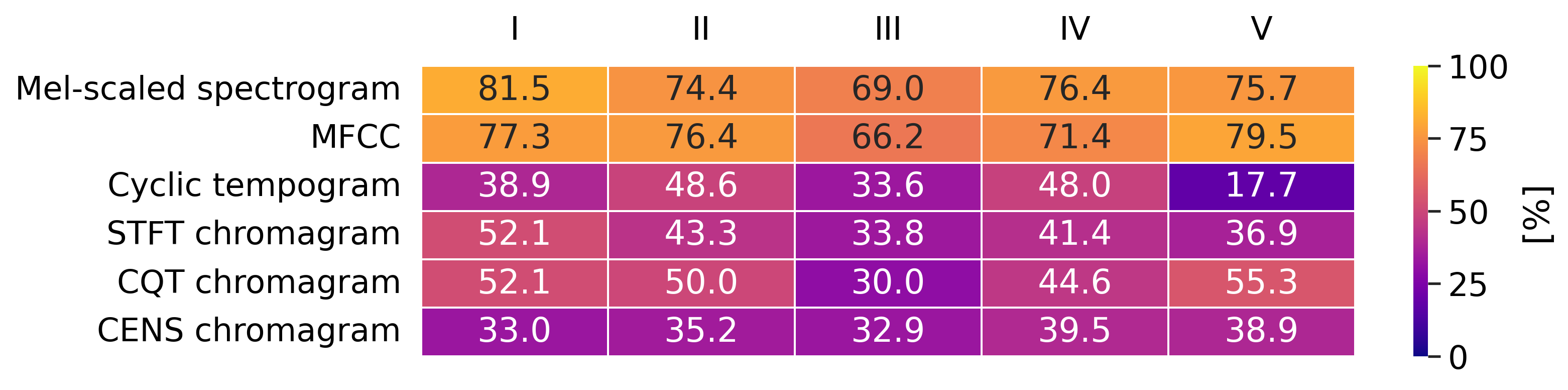}
\caption{Heatmap for F\textsubscript{1} scores in [\%] on audio category level (I~-~V) for six spectral and rhythm features}
\label{fig:heat_f1_score_cat_A}
\end{figure}

\begin{figure}
\includegraphics[width=\textwidth]{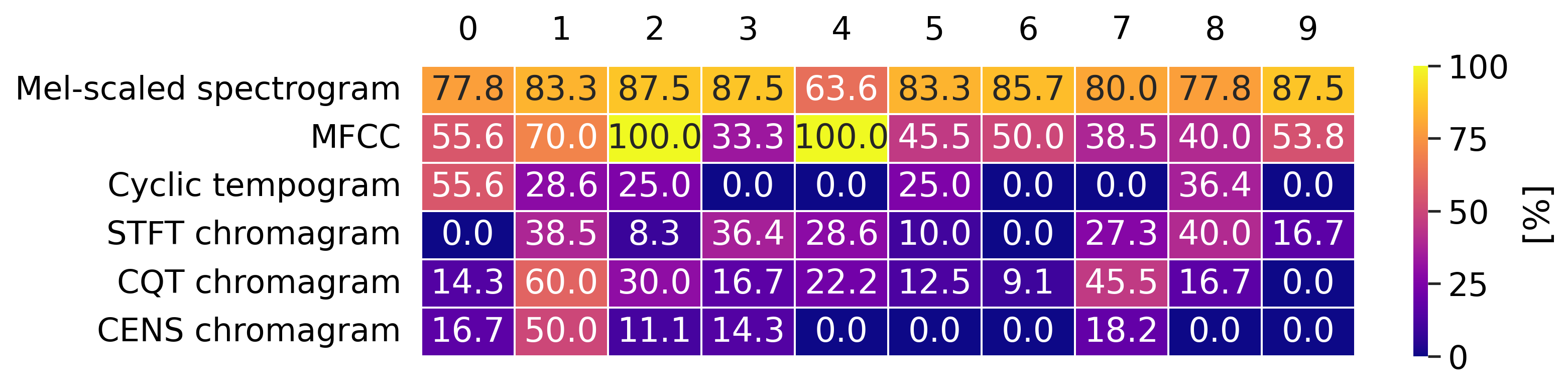}
\caption{Precisions heatmap in [\%] on audio class level (category I, classes 0~-~9) for six spectral and rhythm features for audio classification}
\label{fig:heat_precision_cat_1}
\end{figure}

\begin{figure}
\includegraphics[width=\textwidth]{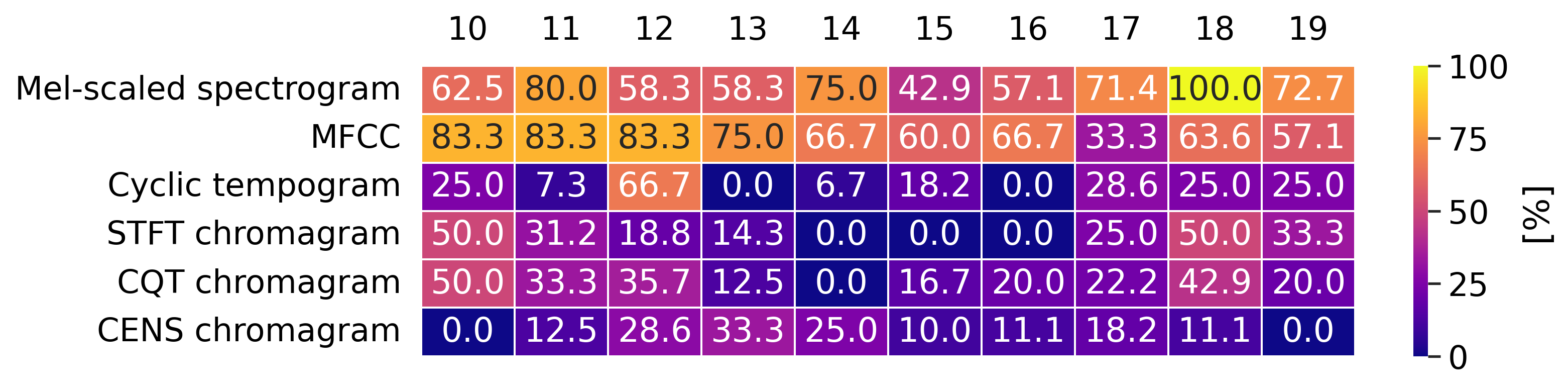}
\caption{Precisions heatmap in [\%] on audio class level (category II, classes 10~-~19) for six spectral and rhythm features for audio classification}
\label{fig:heat_precision_cat_2}
\end{figure}

\begin{figure}
\includegraphics[width=\textwidth]{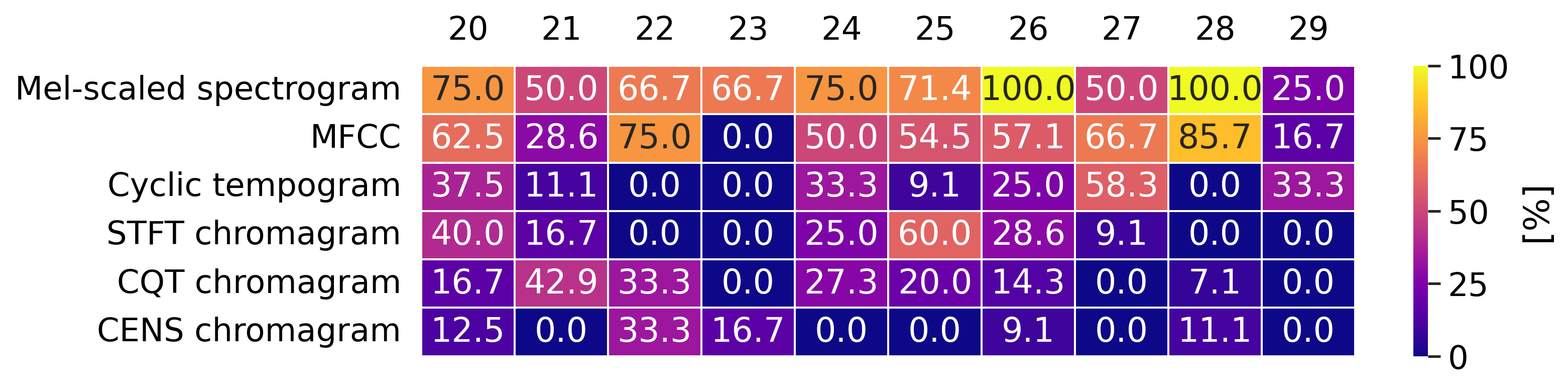}
\caption{Precisions heatmap in [\%] on audio class level (category III, classes 20~-~29) for six spectral and rhythm features for audio classification}
\label{fig:heat_precision_cat_3}
\end{figure}

\begin{figure}
\includegraphics[width=\textwidth]{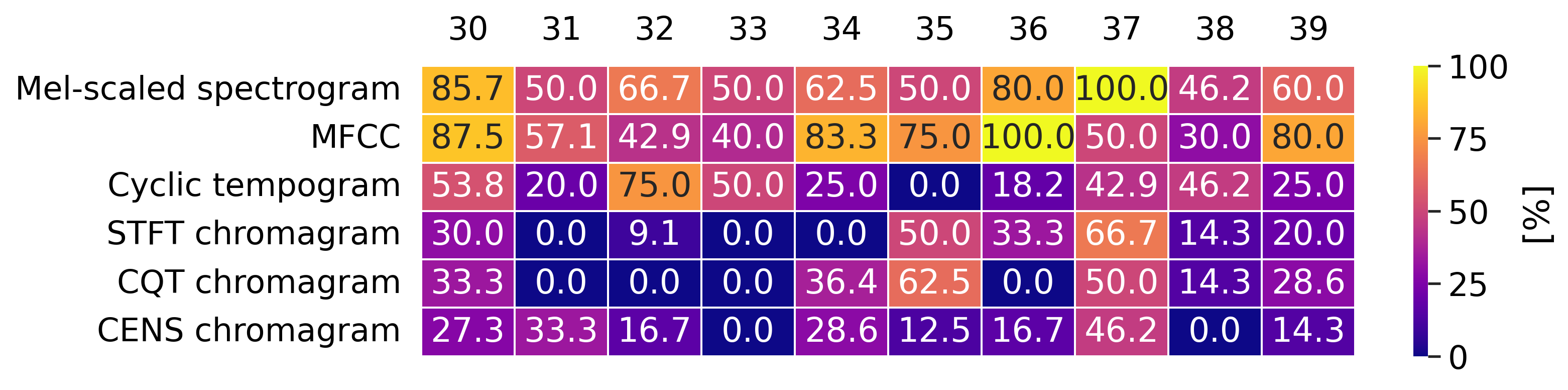}
\caption{Precisions heatmap in [\%] on audio class level (category IV, classes 30~-~39) for six spectral and rhythm features for audio classification}
\label{fig:heat_precision_cat_4}
\end{figure}

\begin{figure}
\includegraphics[width=\textwidth]{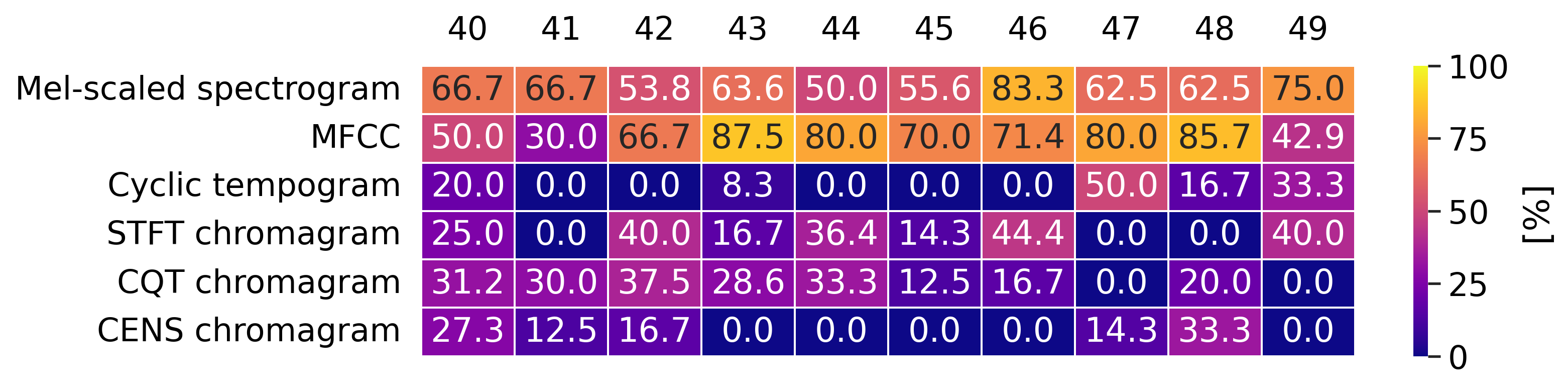}
\caption{Precisions heatmap in [\%] on audio class level (category V, classes 40~-~49) for six spectral and rhythm features for audio classification}
\label{fig:heat_precision_cat_5}
\end{figure}

In \Cref{fig:heat_accuracy_cat_A} to \Cref{fig:heat_f1_score_cat_A} the accuracies, precisions, recalls and F\textsubscript{1} scores calculated in [\%] using the various spectral and rhythm features on audio category level are shown per audio category (I~-~V).
The accuracies (\Cref{fig:heat_accuracy_cat_A}) are calculated per audio category using \Cref{eqn:accuracy}, the precisions (\Cref{fig:heat_precision_cat_A}) are calculated per audio category using \Cref{eqn:precision}, the recall values (\Cref{fig:heat_recall_cat_A}) are calculated per audio category using \Cref{eqn:recall} and the F\textsubscript{1} scores (\Cref{fig:heat_f1_score_cat_A}) are calculated per audio category using \Cref{eqn:f1_score}.

In \Cref{fig:heat_precision_cat_1} to \Cref{fig:heat_precision_cat_5} the precisions calculated in [\%] using the various spectral and rhythm features are shown on audio class level.
The reasons why the precision metric is shown in \Cref{fig:heat_precision_cat_1} to \Cref{fig:heat_precision_cat_5} (multiclass classification on audio class level) instead of accuracy, recall or F\textsubscript{1} score are the following.
Precision is crucial when minimizing false positives is more important than reducing false negatives.
High precision ensures that when a model predicts a certain class, it is more likely to be correct.
Precision is valuable when class-specific confidence is important, where it ensures that when the model predicts a particular class, it does so with high confidence, reducing the risk of misclassifications.
Additionally, in multiclass settings with skewed decision impact, optimizing for precision in critical classes might be preferable.
Finally, precision is often more interpretable for decision-making, as it provides a clear understanding of how reliable the model’s positive predictions are \cite{bishop2006pattern, Murphy2012, Han2012}.

The precisions are calculated using \Cref{eqn:precision}. \Cref{fig:heat_precision_cat_1} shows the results for audio category I (classes 0~-~9), \Cref{fig:heat_precision_cat_2} for audio category II (classes 10~-~19), \Cref{fig:heat_precision_cat_3} for audio category III (classes 20~-~29), \Cref{fig:heat_precision_cat_4} for audio category IV (classes 30~-~39) and \Cref{fig:heat_precision_cat_5} for audio category V (classes 40~-~49).

\section{Discussion}
\label{sec4}

The experimental results presented in \Cref{sec3} (\Cref{fig:heat_accuracy_cat_A} to \Cref{fig:heat_precision_cat_5}) provide comprehensive insights into the performance of the various spectral and rhythm features in multiclass classification tasks across different audio categories and classes.
Overall, the experimental results show clearly that on audio category level the mel-scaled spectrogram and the mel-frequency cepstral coefficients (MFCC) consistently show the best performance in terms of the audio classification capability using the evaluation metrics accuracy, precision, recall and F\textsubscript{1} score across all audio categories I - V.
In 100~\% of the cases the mel-scaled spectrogram and MFCC outperform both the cyclic tempogram as well as the STFT, CQT and CENS chromagram.
This means that spectral features such as mel-scaled spectrogram and MFCC significantly outperform rhythm features like the cyclic tempogram and chromagrams. 

Specifically, for accuracy (\Cref{fig:heat_accuracy_cat_A}) the mel-scaled spectrogram achieves an arithmetic mean of 76.5~\% over the audio categories I - V, while MFCC follows closely at 76.3~\%.
The corresponding values for the tempogram and the chromagrams range between 36.2~\% and 45.6~\%.

For precision (\Cref{fig:heat_precision_cat_A}) the mel-scaled spectrogram achieves an arithmetic mean of 77.7~\% and MFCC 74.4~\% over the audio categories I - V. Tempogram and chromagrams range between 36.3~\% and 46.5~\%.

In terms of recall (\Cref{fig:heat_recall_cat_A}), the mel-scaled spectrogram scores with an arithmetic mean of 75.5~\%, with MFCC at 74.3~\% over the audio categories I - V. The tempogram and chromagram range is between 35.7~\% and 46.5~\%.

The F\textsubscript{1} score (\Cref{fig:heat_f1_score_cat_A}) results are similar to the recall results, with the mel-scaled spectrogram at 75.4~\% and MFCC at 74.2~\%, tempogram and chromagrams ranging between 35.9~\% and 46.4~\%.

These results indicate that mel-scaled spectrogram and MFCC are the most effective feature sets for classification tasks on audio category level compared to tempogram and chromagrams.
It suggests that the mel-scaled spectrogram has the highest potential to capture relevant frequency-based information for differentiating audio classes and MFCC exhibits the second-best performance.

All metrics (accuracy, precision, recall and F\textsubscript{1} score) are approximately 35~\% higher for mel-scaled spectrogram and MFCC compared to tempogram and chromagrams.
For all metrics the values are slightly higher for the mel-scaled spectrogram compared to MFCC.
The largest difference is seen for the precision results where the value for the mel-scaled spectrogram is 3.3~\%  greater than the MFCC value.

On the audio class level we can see a similar trend.
However, at the audio class level (0 - 49) the classification performance is significantly lower compared to the broader category level (I - V) across the board.
Despite this decline, mel-scaled spectrogram and MFCC continue to achieve the best results, although absolute values are lower compared to the values on audio category level.
Since we have seen similar trends for the four metrics accuracy, precision, recall and F\textsubscript{1} score on the audio category level, we focus only on the precision metric for the audio class level.
The reasons why the precision metric is preferred over the other metrics accuracy, recall and F\textsubscript{1} score for audio classification tasks are given in \Cref{sec3}.
Therefore, the results for precision are shown in \Cref{fig:heat_precision_cat_1} to \Cref{fig:heat_precision_cat_5}.
However, similar trends can be found for accuracy, recall and F\textsubscript{1} score.

For the mel-scaled spectrogram the arithmetic mean for precisions over all audio classes 0 - 49 is 69.3~\% and for the MFCC 61.3~\%.
The arithmetic means for precisions for the individual audio categories I - V range from 64.0~\% to 81.4~\% for the mel-scaled spectrogram and from 49.7~\% to 67.2~\% for MFCC.

In contrast, for the cyclic tempogram the arithmetic mean for precisions over all audio classes is 21.3~\% and for the chromagrams 18.4~\%.
The arithmetic means for precisions for the individual audio categories range from 12.8~\% to 35.6~\% for the cyclic tempogram and from 8.3~\% to 25.3~\% for the chromagrams.
This means that the cyclic tempogram offers only limited classification ability, as its performance varies significantly across audio classes.
Chromagram-based features (STFT, CQT and CENS) have a low performance level with low mean precisions, indicating that they are not as useful for this audio classification scenario.

On average the precisions for the mel-scaled spectrogram and MFCC are approximately 46~\% higher compared to tempogram and chromagrams.
When we compare mel-scaled spectrogram and MFCC, we can see that the mel-scaled spectrogram give better precision results compared to MFCC.
The precisions for the mel-scaled spectrogram are approximately 8~\% higher compared to MFCC.
In summary this shows that the mel-scaled spectrogram and MFCC perform also much better on audio class level compared to the cyclic tempogram and the STFT, CQT and CENS chromagrams.
The mel-scaled spectrogram shows the highest average precision (69.3~\%) which makes it the most effective feature for audio classification on audio class level. 

The analysis of the precision results from \Cref{fig:heat_precision_cat_1} to \Cref{fig:heat_precision_cat_5} shows that in most cases the mel-scaled spectrogram and MFCC show higher precision values as compared to the tempogram and chromagrams.
However, there are a few exceptions on audio class level in which the mel-scaled spectrogram and/or the MFCC are/is outperformed by the cyclic tempogram or the STFT, CQT or CENS chromagrams.
Some audio classes show poor results even with MFCC (like for example \Cref{fig:heat_precision_cat_3}, audio classes 23 'breathing' and 29 'drinking, sipping').
These audio classes are challenging to correctly classify due to several reasons.
First, both audio classes are composed of soft, low-intensity sounds with the spectral energy distributed in a low-frequency range.
Additionally, these audio classes have a low signal-to-noise ratio (SNR) which means that their low amplitude makes them more susceptible to background noise, while MFCC, which emphasize spectral envelope characteristics, may fail to capture subtle variations.
Another limitation comes from MFCC discarding phase information which leads to a loss of crucial temporal structures.
Furthermore, both audio classes exhibit high intra-class variability.
'Breathing' can vary based on factors like speed and depth, while 'drinking, sipping' can differ depending on liquid type, container sounds or swallowing.
This increases the difficulty of learning consistent features.

On the technical side there are the following reasons why some of the spectral and rhythm features perform better or worse compared to others for audio signals from the ESC-50 dataset on audio category level respectively audio class level.
The mel-scaled spectrogram captures both temporal and spectral characteristics which makes it ideal for speech and music classification, though it is sensitive to noise.
MFCCs reduce dimensionality while preserving key spectral properties but may not retain fine-grained temporal details.
The cyclic tempogram highlights tempo periodicities which are useful for rhythm-based tasks but ignores absolute tempo.
STFT chromagrams encode harmonic structures for chord and tonality classification but are sensitive to noise and have limited resolution.
CQT chromagrams improve frequency resolution at lower pitches, aiding melody and chord recognition, though they are computationally expensive.
CENS chromagrams provide stable harmonic features robust to dynamic variations but lack fine spectral details.

\section{Conclusion}
\label{sec5}

The results of this research provide a detailed analysis of the performance of various spectral- and rhythm-based features in the context of multiclass audio classification with deep CNNs on the ESC-50 dataset.
The findings demonstrate that spectral features, particularly the mel-scaled spectrogram and the mel-frequency cepstral coefficients (MFCC), outperform rhythm features such as the cyclic tempogram and chromagrams (STFT, CQT and CENS) across all tested audio categories and classes.
In conclusion, mel-scaled spectrogram and MFCC are the most reliable features for audio classification tasks on the ESC-50 dataset using a deep CNN, whearas rhythm-based features such as the cyclic tempogram and chromagrams are less effective.

On audio category level with mel-scaled spectrograms and MFCCs accuracies of 76.5~\% / 76.3~\% , precisions of 77.7~\% / 74.4~\%, recall values of 75.5~\% / 74.3~\% and F\textsubscript{1} scores of 75.4~\% / 74.2~\% can be achieved (arithmetic means across all audio categories I - V).
The other spectral and rhythm features (cyclic tempograms, short-time Fourier transform (STFT) chromagrams, constant-Q transform (CQT) chromagrams and chroma energy normalized statistics (CENS) chromagrams) show lower values for accuracy, precision, recall and F\textsubscript{1} score.
Tempograms and chromagrams are ranging between 36.2~\% and 45.6~\% for acuracy, 36.3~\% and 46.5~\% for precision, 35.7~\% and 46.5~\% for recall and 35.9~\% and 46.4~\% for F\textsubscript{1} score.
These values are also arithmetic means over all audio categories I - V.
In general the audio classification performance for all spectral and rhythm features is higher on audio category level compared to audio class level.

While the current study was conducted using the ESC-50 dataset \cite{Piczak2015ESCDF}, future work could extend the analysis to other environmental sound classification datasets such as UrbanSound8K \cite{salamon2014dataset}, AudioSet \cite{gemmeke2017audio} or DESED \cite{Turpault2019}.
These datasets provide a diverse range of audio recordings that can help validate the generalizability of the findings.
Additionally, the methodologies explored in this research have potential applications in various fields, including speech recognition \cite{Deng2013}, music genre classification \cite{bergstra2006aggregate}, bioacoustics \cite{Stowell2019} and environmental monitoring \cite{Mitilineos2018TwoLevel}.
For instance, MFCC and mel-scaled spectrograms have already shown promise in areas such as speaker identification \cite{Reynolds2000SpeakerIV}, automatic music tagging \cite{choi2018automatic} and machine listening applications for smart cities \cite{isler2025urban}.

Despite the strong performance of spectral features, there are several limitations to this study. First, the ESC-50 dataset, while widely used, consists of only 2,000 audio samples, which may not be sufficiently large to generalize results to real-world applications.
Second, certain classes, such as ‘breathing’ and ‘drinking, sipping,’ exhibit inherent challenges due to their low signal-to-noise ratio (SNR) and high intra-class variability.
This indicates that classification accuracy may not be solely dependent on the choice of feature set but also on dataset quality and balance.

The balance of samples within the ESC-50 dataset plays a crucial role in classification performance.
Although the dataset is designed to have an equal number of samples across 50 classes, certain audio categories inherently present greater classification difficulty due to overlapping spectral properties or low signal-to-noise ratio (SNR).
This imbalance in feature distinguishability, rather than raw sample count, affects the ability of some classifiers to generalize effectively.
Future research could explore data augmentation techniques or synthetic data generation to mitigate these effects.

The observed performance gap between the different spectral and rhythm features is primarily attributed to their inherent ability to capture distinguishing audio characteristics.
The mel-scaled spectrogram and MFCC perform very good due to their capacity to represent frequency-based structures which are crucial for human auditory perception.
Conversely, rhythm features like the cyclic tempogram are designed to capture periodicities, making them more suitable for music or rhythm-based classification rather than general environmental sound recognition.
The lower classification performance of chromagram-based features can be explained by their emphasis on harmonic structures, which may not be as relevant in a dataset like ESC-50, which contains a broad range of non-harmonic environmental sounds.

The ESC-50 dataset is a benchmark for audio classification tasks and many papers present innovative approaches to achieve high performance levels.
Although it is not the aim of this research work to achieve a further improvement in classification performance in relation to the ESC-50 dataset, a brief overview of the most successful approaches in relation to classification performance for the ESC-50 dataset is presented here.
Among the most successful papers, Piczak (2015) \cite{Piczak2015} introduced a convolutional neural network (CNN) architecture that achieved an accuracy of 81.3~\%.
Later, Piciarelli et al. (2017) \cite{piciarelli2017fusion} proposed a fusion of hand-crafted features and deep learning, reaching an accuracy of 83.5~\%.
The work by Sainath et al. (2018) \cite{sainath2018low} employed a transfer learning approach using a pre-trained CNN, achieving an accuracy of 85.1~\%.
In 2019, Guo et al. \cite{guo2019multi} presented a multi-resolution CNN framework that obtained an accuracy of 86.5~\%.
Another notable work by Zhang et al. (2020) \cite{zhang2020graph} utilized a graph convolutional network (GCN) to model audio patterns, achieving an accuracy of 87.2~\%.
Most recently, Li et al. (2022) \cite{li2022hybrid} proposed a hybrid approach combining CNN and GCN, which achieved the highest accuracy of 88.5~\% on the ESC-50 dataset. 

Several areas for future research can be derived from the work presented in this paper.
The first area is deep learning-based feature extraction where spectral and rhythm features can be combined to explore CNN-based feature extraction and hybrid approaches. 
Secondly, by multi-modal audio analysis additional features such as wavelet transforms or auditory filterbanks can be integrated for a more comprehensive analysis.
The third area of future research is the  dataset expansion and generalization where the methods are tested on larger and more diverse datasets like AudioSet \cite{gemmeke2017audio} and FSD50K \cite{Fonseca2022FSD50K}.
Additionally, the robustness to noise plays a crucial role and therefore 
noise reduction techniques to improve classification performance on low signal-to-noise ratio (low-SNR) classes can be investigated.
Finally, it is important to focus on real-world applications to be able to implement real-time classification systems for applications in healthcare (e.g. respiratory sound analysis), security (e.g. anomaly detection via sound monitoring) or smart environments.

In summary, the research presented in this paper highlights the superiority of spectral features for multiclass environmental sound classification, with mel-scaled spectrogram and MFCC emerging as the most effective features using a deep CNN for the ESC-50 dataset.
Rhythm-based features like the cyclic tempogram and chromagrams are less suited for general sound classification tasks.
The findings lay a foundation for further research, particularly in deep learning-driven approaches and real-world applications to ensure continued advancements in machine listening and audio signal processing.

\end{document}